\documentclass[onecolumn,journal,draftclsnofoot,12pt]{IEEEtran}

\usepackage{bm}
\usepackage{amssymb}
\usepackage{amsmath}
\usepackage{amsfonts}
\usepackage{graphicx}
\usepackage{subfigure}
\usepackage{cite}
\usepackage{enumerate}
\usepackage{url}
\usepackage{epstopdf}
\usepackage{float}
\usepackage{color}
\usepackage{setspace}
\usepackage{lineno}
\usepackage[utf8x]{inputenc}
\usepackage{enumitem}

\newcommand{\beq}{\begin{equation}}
\newcommand{\eeq}{\end{equation}}

\allowdisplaybreaks[2]

\makeatletter  
\newif\if@restonecol  
\makeatother

\usepackage[linesnumbered,ruled,vlined]{algorithm2e}
\usepackage{algpseudocode}  
\usepackage{amsmath}  

\usepackage{makecell} 

\renewcommand{\Pr}{\mathrm{Pr}}

\DeclareMathOperator*{\argmax}{argmax}

\newtheorem{theorem}{\bf Theorem}

\newcounter{optcnt}
\renewcommand\theoptcnt{\arabic{optcnt}}

\begin{document}

\title{\setstretch{1.4}\huge{Cooperative Internet of UAVs: Distributed Trajectory Design by Multi-agent Deep Reinforcement Learning}}

\author{
\IEEEauthorblockN{
\normalsize{Jingzhi~Hu},~\IEEEmembership{\normalsize Student~Member,~IEEE},
\normalsize{Hongliang~Zhang},~\IEEEmembership{\normalsize Member,~IEEE},
\normalsize{Lingyang~Song},~\IEEEmembership{\normalsize Fellow,~IEEE},
\normalsize{Robert~Schober},~\IEEEmembership{\normalsize Fellow,~IEEE},
and~\normalsize{H.~Vincent~Poor},~\IEEEmembership{\normalsize Fellow,~IEEE}\\
}

\thanks{This work has been presented in part at IEEE Globecom 2019 \cite{mypaper}.}
\thanks{
 J. Hu, and L. Song are with Department of Electronics, Peking University, Beijing, China (email: \{jingzhi.hu, lingyang.song\}@pku.edu.cn).
 }
 \thanks{
 H. Zhang is with Department of Electronics, Peking University, Beijing, China, and also with Department of Electrical Engineering, Princeton University, Princeton, NJ, USA (email: hongliang.zhang92@gmail.com)
 }
  \thanks{
 R.~Schober is with Institute of Digital Communications, Friedrich-Alexander University of Erlangen-Nuremberg, Erlangen, Germany (email: robert.schober@fau.de).
}
 \thanks{
 H.~V.~Poor is with Department of Electrical Engineering, Princeton University, Princeton, NJ, USA~(email: poor@princeton.edu).
 }
}

\maketitle
\vspace{-3.4em}
\begin{abstract}
Due to the advantages of flexible deployment and extensive coverage, unmanned aerial vehicles~(UAVs) have great potential for sensing applications in the next generation of cellular networks, which will give rise to a cellular Internet of UAVs. In this paper, we consider a cellular Internet of UAVs, where the UAVs execute sensing tasks through cooperative sensing and transmission to minimize the age of information~(AoI). However, the cooperative sensing and transmission is tightly coupled with the UAVs' trajectories, which makes the trajectory design challenging. To tackle this challenge, we propose a distributed sense-and-send protocol, where the UAVs determine the trajectories by selecting from a discrete set of tasks and a continuous set of locations for sensing and transmission. Based on this protocol, we formulate the trajectory design problem for AoI minimization and propose a compound-action actor-critic~(CA2C) algorithm to solve it based on deep reinforcement learning. The CA2C algorithm can learn the optimal policies for actions involving both continuous and discrete variables and is suited for the trajectory design.  {Our simulation results show that the CA2C algorithm outperforms four baseline algorithms}. Also, we show that by dividing the tasks, cooperative UAVs can achieve a lower AoI compared to non-cooperative UAVs.
\end{abstract}


\newpage

\setlength{\abovedisplayskip}{3pt}
\setlength{\belowdisplayskip}{3pt}

\section{Introduction}
\label{sec: introduction}
Unmanned aerial vehicles (UAVs) are an emerging technology that has been effectively applied in military, public, and civil applications \cite{gupta2015survey}.
Due to advantages such as on-demand flexible deployment and extensive service coverage \cite{hayat2016survey}, UAVs have been widely used in critical sensing applications, where they need to execute multiple sensing tasks and transmit the results to base stations (BSs) over cellular networks~\cite{Zhang2020Beyond}.
This general concept is referred to as a cellular Internet of UAVs~\cite{Zhang2019Cooperation, Zhang2019Cellular}.
For certain sensing applications, such as traffic monitoring~\cite{puri2007statistical}, collapsed building detection \cite{yang2018real}, and forest fire surveillance \cite{casbeer2006cooperative}, UAVs need to continuously sense and transmit the results to BSs, in order to keep the sensing results as fresh as possible.

To quantify the freshness of sensing results, the age of information~(AoI) has been introduced~\cite{kaul2011minimizing}.
To be specific, for sensing applications, the AoI of a task is defined as the time that has elapsed since the most recent successful transmission of a sensing result~\cite{Kaul2012Real}.
Then, for a given time duration, the freshness of the sensing results in the system can be evaluated by the accumulated AoI of the tasks.
For the applications considered in this paper, the tasks are scattered across different locations, and the sensors have limited sensing range.
Hence, the UAVs need to approach the locations of the tasks in order to sense successfully, which may result in long flight durations.
Therefore, in order to minimize the AoI, it is advantageous for the UAVs to perform cooperative sensing, as this can reduce the flight duration.

In this paper, we consider the trajectory design problem in a cellular Internet of UAVs, where the UAVs continuously execute multiple sensing tasks through cooperative sensing and transmission, with the objective to minimize the accumulated AoI of the tasks.
As a sensing task can be carried out by multiple cooperative UAVs and the number of possible trajectories of each UAV is indefinite, a centralized trajectory design deployment can lead to high computational complexity, and thus, the UAVs should design their trajectories in a distributed manner.
However, due to the cooperation, the trajectories of different UAVs will influence each other.
Moreover, each UAV needs to jointly consider the probability of successful sensing and the transmission rate when designing its trajectory, since a lower probability of successful sensing and a lower data rate will result in a higher AoI.
Therefore, the distributed trajectory design for the cooperative Internet of UAVs is a challenging problem.

To handle these challenges, we propose a distributed sense-and-send protocol, which facilitates the cooperation of the UAVs. 
Based on this protocol, the trajectory design for the Internet of UAVs can be modeled by a Markov decision process~(MDP), where the UAVs aim to minimize the accumulated AoI of the tasks.
In the MDP, the UAVs design their trajectories in a distributed manner and optimize their trajectory design policies by learning from previous design experiences.
Since the UAVs have to cope with uncertain environments with high-dimensional state spaces, we adopt deep reinforcement learning approaches in this paper.
Specifically, we propose a compounded-action actor-critic~(CA2C) algorithm to determine the optimal trajectory design policies of the UAVs in an efficient manner.

In the literature, most related works focus only on UAV sensing problems, but do not consider sensing and transmission jointly.
In \cite{tisdale2009autonomous}, the autonomous path planning problem was discussed for a swarm of UAVs equipped with vision-based sensing systems to search for a stationary target.
In \cite{mersheeva2015multi}, the authors proposed a heuristic algorithm for multiple UAVs to continuously monitor multiple regions to maximize the visiting frequency of each region.
In \cite{stump2011multi}, the authors formulated the trajectory design problem for cooperative UAV sensing applications as a routing problem with time windows and solved it using exact methods.
In \cite{semsch2009autonomous}, the authors proposed an occlusion-aware surveillance algorithm to address trajectory design in the presence of realistic occlusion and motion constraints.

In the few works that have considered both sensing and transmission for the cellular Internet of UAVs, the UAVs either do not cooperate~\cite{hu2018reinforcement} or cooperate in a centralized manner~\cite{scherer2016persistent},~\cite{zhang2018cellular}.
In~\cite{hu2018reinforcement}, the authors considered multiple non-cooperative UAVs where each UAV had a different sensing task, and proposed a multi-agent reinforcement learning algorithm for trajectory design.
However, the cooperation of the UAVs was not discussed in~\cite{hu2018reinforcement}.
In \cite{scherer2016persistent}, the authors proposed a centralized offline path planning algorithm for multi-UAV surveillance under energy and communication constraints.
In \cite{zhang2018cellular}, an iterative trajectory, sensing, and scheduling algorithm was proposed for the centralized design of the UAVs' trajectories.
However, the distributed trajectory design for a cooperative Internet of UAVs has not been considered, yet.
Besides,~\cite{hu2018reinforcement,scherer2016persistent,zhang2018cellular} and other related works have not studied continuous sensing for AoI minimization.

The main contributions of this paper can be summarized as follows:
\begin{itemize}
    \item We propose a distributed sense-and-send protocol to coordinate sensing and transmission for execution of multiple sensing tasks in the cellular Internet of UAVs.
    \item  {We formulate the trajectory design problem for the Internet of UAVs as an MDP with the objective to minimize the accumulated AoI for a given duration. We also develop a deep reinforcement learning algorithm, which we refer to as CA2C algorithm, to solve the MDP.}
    \item  {Our simulation results show that the proposed CA2C algorithm outperforms the deep deterministic policy gradient~(DDPG) algorithm, the deep Q-network~(DQN) algorithm, and two baseline conventional~(i.e., non-learning based) algorithms.} Besides, our results also confirm the performance gain achieved with cooperative UAVs in terms of AoI minimization.
\end{itemize}

The rest of this paper is organized as follows.
In Section~II, the system model for the cellular Internet of UAVs is provided. 
In Section~III, we propose a distributed sense-and-send protocol for coordinating the UAVs for sensing and transmission of the sensing results.
In Section~IV, the trajectory design problem for the cooperative Internet of UAVs is formulated as an MDP with the objective to minimize the accumulated AoI.
In Section~V, the CA2C algorithm for solving the MDP is derived.
In Section~VI, we analyze the convergence and complexity of the proposed CA2C algorithm as well as the optimal flying altitude of the UAVs.
Simulation results are presented in Section~VII, and conclusions are drawn in Section~VIII.

\section{System Model}
\label{sec: system model}

\begin{figure}[!t]
\center{\includegraphics[width=0.6\linewidth]{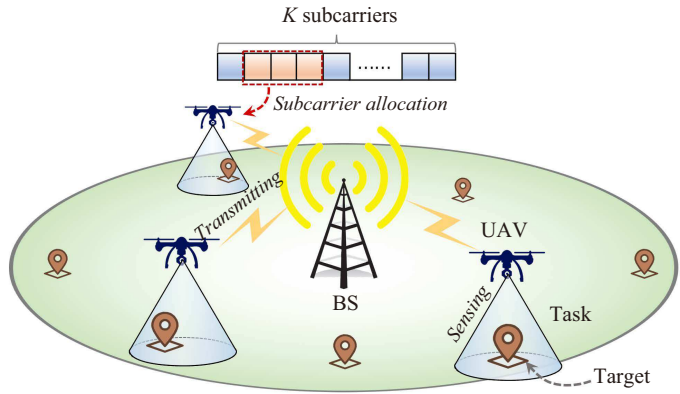}}
	\caption{ Cellular Internet of UAVs.}
	\label{fig: system model}
	\vspace{-1em}
\end{figure}

We consider a single-cell cellular Internet of UAVs as shown in Fig.~\ref{fig: system model}.
 {The BS is located at the center of the cell and employs $K$ subcarriers with bandwidth $W$ for uplink transmission.}
The coverage area of the BS is a circular region with radius $R_c$.
Within the cell coverage, there is a swarm of UAVs, $\mathcal M = \{1,2,...,M\}$, which are indexed by $i$\footnote{
    In this paper, we assume that the UAVs adopt proper collision avoidance mechanisms which prevent them from colliding with each other. The design of collision avoidance mechanisms is beyond the scope of this paper
}.
The number of subcarriers,~$K$, is assumed to be larger than the number of UAVs, $M$.
The swarm of UAVs continuously and cooperatively executes a set of sensing tasks, $\mathcal N=\{1,2,...,N\}$, which are indexed by $j$.
Each Task $j$ involves a sensing target, which is located at $\bm x^t_j = (x^t_j,y^t_j,0)$, to be sensed by the UAVs.
We denote the location of the BS by $\bm x^b= (0,0,H_0)$, and the location of UAV $i$ by $\bm x_i = (x_i,y_i,h_i)$.
All UAVs are assumed to be flying at the same altitude, i.e., $h_i = h,~\forall i \in \mathcal M$.
	
The UAVs execute the sensing tasks in two steps: UAV sensing and UAV transmission, where the UAVs first  {sense the targets}, and then transmit the results back to the BS~\cite{Hu2020Reinforcement}.
To keep the sensing results received by the BS as fresh as possible, the UAVs sense and transmit continuously in a cooperative manner.
The freshness of the a sensing result is evaluated by its \emph{AoI}.
Specifically, the cooperation is achieved by allowing the UAVs to dynamically divide the tasks into different sets, and each UAV executes the tasks in one set. 
Which tasks a UAV has to execute will significantly affect its trajectory design.

\subsection{UAV Sensing}
The UAVs are equipped with onboard sensors to sense the targets.
However, due to the practical limitations of the sensing range and precision, the sensing is not always successful \cite{Hossain2012Impact}.
In order to perform successful sensing, a UAV needs to ensure that the sensing target is within its maximum sensing angle $\phi$~\cite{zhang2018joint}.
To evaluate the probability of successful sensing, we adopt a probabilistic sensing model.
To be specific, given that the target at $\bm x^t_j$ is within the maximum sensing angle of UAV $i$ located at $\bm x_i$, the probability of successful sensing is modeled as
\beq
p_s(\bm x_i, \bm x^t_j) = e^{-\lambda \|\bm x_i - \bm x^t_j\|_2},
\eeq
where $\lambda$ is a parameter that depends on the quality of the sensor~\cite{Shakhov2017Experiment},
$\|\cdot\|_2$ denotes the Euclidean norm, and $\|\bm x_i - \bm x^t_j\|_2$ is the distance between $\bm x_i$ and $\bm x^t_j$.
Therefore, based on~\cite{zhang2018joint},~\cite{Shakhov2017Experiment}, the probability of successful sensing for UAV $i$ at $\bm x_i$ with respect to the target of Task $j$ is
\beq
\label{equ: successful task-sensing probability}
{p}_{s,j}(\bm x_i)=
\begin{cases}
	p_s(\bm x_i, \bm x^t_j), &\text{if }\|\bm x_i-\bm x^t_j\|_2\cdot \sin\phi\leq r_s, \\
	0, & \text{otherwise}.
\end{cases}
\eeq

For UAV $i$ located at $\bm x_i = (x_i,y_i,h)$, its \emph{sensing range} can be modeled as a circular region on the ground with radius $r_s = h\cdot\tan\phi$ centered at $(x_i,y_i,0)$.
Based on~(\ref{equ: successful task-sensing probability}), if the target of a task is within the sensing range of a UAV, the UAV will have non-zero probability of successful sensing with respect to the task; otherwise, the probability for the UAV to sense the target of the task successfully will be $0$.
If the sensing was successful, we refer to the corresponding sensing result as a \emph{valid result}, otherwise, we refer to the sensing result as an \emph{invalid result}.
Besides, we assume that the amount of data needed to represent the sensing result is $D_s$ bits.

Moreover, we assume that the UAVs cannot determine whether the sensing was successful or not based on their sensing results, because of their limited data processing capabilities.
Therefore, the UAVs need to send their sensing results to the BS, and the BS will decide whether the sensing results are valid or not.

\subsection{UAV Transmission}
\label{sec: UAV transmission}
During UAV transmission, the UAVs transmit their sensing results to the BS over orthogonal uplink channels to avoid interference.
Denote the transmit power of the UAVs by $P$.
Then, the received signal-to-noise ratio (SNR) at the BS for the UAV at $\bm x$ can be expressed as
\beq \label{Equ: received power of BS}
\gamma(\bm x)= \frac{P}{N_010^{\mathrm{PL}_{a}(\bm x)/10}},
\eeq
where $\mathrm{PL}_{a}(\bm x)$ [dB] denotes the average air-to-ground pathloss, and $N_0$ denotes the noise power at the receiver of the BS~\cite{Zhang2018JointRelay}.
To be specific, the pathloss $\mathrm{PL}_{a}(\bm x)$ is averaged over two cases, namely, line-of-sight (LoS) and non-LoS (NLoS) channels~\cite{zhang2018joint, Athukoralage2016Regret}.
Therefore, the average air-to-ground pathloss from the UAV at $\bm x$ to the BS is given by 
\beq
\label{equ: pathloss}
\mathrm{PL}_{a}(\bm x) =
 	\Pr_{\mathrm{LoS}}(\bm x)\cdot \mathrm{PL}_{\mathrm{LoS}}(\bm x)
	+ (1-\Pr_{\mathrm{LoS}}(\bm x))\cdot\mathrm{PL}_{\mathrm{NLoS}}(\bm x).
\eeq
Here, $\Pr_{\mathrm{LoS}}(\bm x)$ is the probability that the channel between the UAV at $\bm x$ and the BS is LoS, and
	$\mathrm{PL}_{\mathrm{LoS}}(\bm x)$ and $\mathrm{PL}_{\mathrm{NLoS}}(\bm x)$ denote the pathlosses for the LoS and NLoS channels, respectively.
To be specific, we adopt the 3GPP channel model in~\cite{3GPP2017R15}, where $\Pr_{\mathrm{LoS}}(\bm x)$ is calculated as
\beq
\Pr_{\mathrm{LoS}}(\bm x)=\begin{cases}
	1,& r(\bm x) \leq r_{c},\\
	\frac{r_{c}}{r(\bm x)} + e^{-r(\bm x)/p_{0}+r_{c}/p_{0}}, &r(\bm x) >r_{c},
\end{cases},
\eeq
where $r(\bm x) = \sqrt{x^2+y^2}$, $p_{0}= 233.98\log_{10}(h) - 0.95$,
$r_{c} =\max\{ 294.05\log_{10}(h) - 432.94, 18\}$,
 and the pathlosses $\mathrm{PL}_{\mathrm{LoS}}(\bm x)$ and $\mathrm{PL}_{\mathrm{NLoS}}(\bm x)$ for the LoS and NLoS channels are given by
\begin{align}
	&\mathrm{PL}_{\mathrm{LoS}}(\bm x)\!=\!30.9\!+\!(22.25-0.5\log_{10}(h))\cdot\log_{10}(\|\bm x - \bm x^b\|_2) \nonumber \\
    & \qquad \qquad +20\log_{10}(f_c),\\
	&\mathrm{PL}_{\mathrm{NLoS}}(\bm x)\!=\! 32.4\!+\!(43.2-7.6\log_{10}(h))\cdot\log_{10}(\|\bm x - \bm x^b\|_2) \nonumber \\
    & \qquad \qquad +20\log_{10}(f_c),
\end{align}
respectively. Here, $f_c$ denotes the carrier frequency.

Assuming that $k$ subcarriers are allocated to the UAV at $\bm x$, the uplink data rate of the UAV is given by
\beq
 R(k, \bm x)= k\cdot W\cdot \log_2(1+\gamma(\bm x)),
\eeq
where $W$ denotes the bandwidth of a subcarrier.

\subsection{Age of Information}
\begin{figure}[!t]
\center{\includegraphics[width=0.45\linewidth]{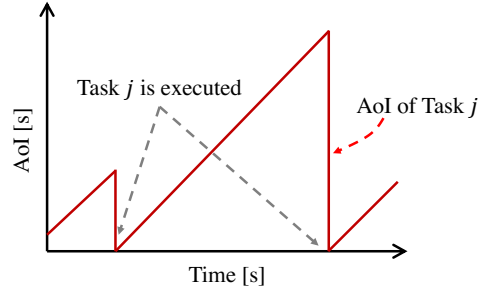}}
	\caption{AoI of Task $j$ versus time.}
	\label{fig: AoI}
	\vspace{-1em}
\end{figure}

To characterize the freshness of the sensing results, the AoI metric is adopted~\cite{Zhang2020Age}.
At time $t$, the AoI of Task $j$ is defined as 
\beq
\label{equ: aoi def}
	 \tilde{\tau}_j = t-U_j,
\eeq
where $U_j$ denotes the last time when Task $j$ was \emph{executed}, i.e., its target was sensed successfully and the transmission of the sensing result was completed.
As shown in Fig.~\ref{fig: AoI}, the AoI of a task is reduced to zero if the task is successfully executed; otherwise, it increases with time.

\section{Distributed Sense-and-Send Protocol}
\label{sec: protocol}

\begin{figure}[!t] 
	\center{\includegraphics[width=0.8\linewidth]{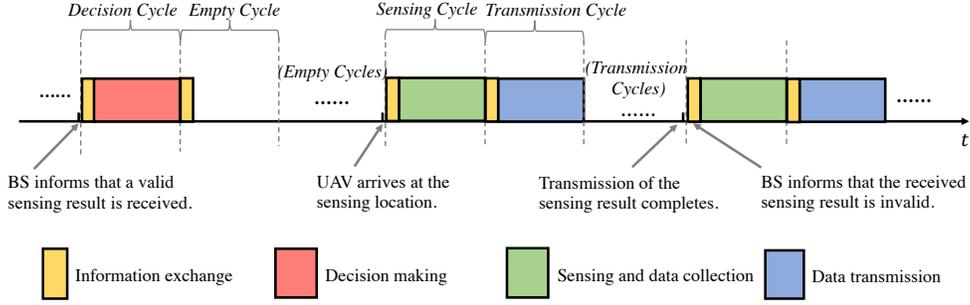}}
	\vspace{-1.8em}
	\caption{ {Distributed sense-and-send protocol.}}
	\label{fig: protocol}
	\vspace{-1.2em}
\end{figure}

In this section, we present a distributed sense-and-send protocol for the cellular Internet of UAVs.
The timeline of the protocol is divided into cycles, and the duration of each cycle is denoted by $t_c$.
A cycle serves as the minimal scheduling unit and can be categorized into four types, i.e., the \emph{decision cycle}, \emph{empty cycle}, \emph{sensing cycle}, and \emph{transmission cycle}, as shown in Fig.~\ref{fig: protocol}.
Each cycle begins with a common information exchange between the UAVs and the BS, and then proceeds to the different actions which distinguish the different types of cycles.

In this protocol, each UAV executes only one task at a time.
For convenience, we define the task which the UAV is executing as its {\emph{selected task}} and define the location where the UAV performs sensing as its \emph{sensing location}.
The UAV will not select a new selected task before it has been informed that a valid sensing result for its selected task has been received by the BS.

In the information exchange, each UAV reports its state to the BS, which includes its current location, selected task, sensing location for that task, and the amount of data from the previous sensing result that is still awaiting transmission.
Then, the BS broadcasts the state of the system, including the states of all the UAVs and the AoI of each task.

The process for a UAV to execute its selected task can be described as follows, see also Fig.~\ref{fig: protocol}.
When a UAV is informed that a valid sensing result for its selected task has been received by the BS, it starts a new \emph{task execution process}.
The task execution process begins with {one} decision cycle, where the UAV determines its next selected task and the corresponding sensing location.
After the decision cycle, the UAV moves towards its new sensing location directly with maximum speed $v_{\max}$.
Until the UAV arrives at the sensing location, it is in an empty cycle.
The number of empty cycles between the decision cycle and the first sensing cycle depends on the distance between the UAV and its sensing location.
After the UAV has arrived at the sensing location, it starts a sensing cycle.
Subsequently, the UAV transmits the sensing result to the BS in the following transmission cycles.
The number of transmission cycles after the sensing cycle is determined by the amount of data generated by the sensing result and the transmission rate.
In the following, we will explain the actions taken in the different types of cycles in detail.

\subsubsection{Decision Cycle}
The UAV reaches the decision cycle as soon as it has completed the sensing and transmission required for its selected task.
During the decision cycle, the UAV determines its new selected task and the corresponding sensing location.
The proposed distributed approach for the UAVs to make their decisions regarding the selected tasks and sensing locations will be introduced in Section~\ref{sec: algorithm design}.
Moreover, during this cycle, we assume that the UAV remains in its current location until it has completed the decision process.

\subsubsection{Empty Cycle}
A UAV transitions to the empty cycle when it has determined its selected task and the corresponding sensing location, but has not arrived at the sensing location yet.
Since a UAV will not select a new selected task before it has successfully executed its current selected task, the UAVs in the empty cycle move with speed $v_{\max}$ directly towards their respective sensing locations.
Suppose UAV $i$ has selected Task $T_i$ and corresponding sensing location $\hat{\bm x}_i$.
Then, the trajectory of the UAV in the empty cycle can be expressed as follows:
\begin{equation}
\label{equ: next location}
	\bm x_{i}' - \bm x_{i}
	= \begin{cases}
 	\hat{\bm x}_i- \bm x_i, \quad \text{if } \|\hat{\bm x}_i - \bm x_i\|_2 \leq v_{\max}t_{c}, \\
 	\frac{\hat{\bm x}_i - \bm x_i}{\|\hat{\bm x}_i - \bm x_i\|_2}\cdot v_{\max}t_{c}, \quad \text{otherwise,}
 \end{cases}
\end{equation}
where $\bm x_i$ and $\bm x_i'$ denote the locations of UAV $i$ at the beginning of the current cycle and at the beginning of the next cycle, respectively.
Here, for the first case in~(\ref{equ: next location}), UAV $i$ will reach its new sensing location in the current cycle as its distance to the sensing location is within the maximum flight distance; 
for the second case in~(\ref{equ: next location}), UAV $i$ moves with the maximum speed towards the sensing location, but will not reach it in the current cycle.

\subsubsection{Sensing Cycle}
The UAVs in the sensing cycle sense their respective targets.
Without loss of generality, we assume that the sensing process can be completed within one cycle, and the UAVs hover over their sensing locations during the sensing cycle.
The probability for UAV $i$ at sensing location $\hat{\bm x}_i$ to sense successfully is $p_{s,T_i}(\hat{\bm x}_i)$, which is given in~(\ref{equ: successful task-sensing probability}).
 {During the sensing cycle, the UAV collects $D_s$ bits of sensing data.}
Since the UAVs cannot determine whether the sensing result is valid or not, they transmit their sensing results to the BS for discrimination.

\subsubsection{Transmission Cycle}
The UAVs that have completed the sensing cycle enter the data transmission cycle, where the UAVs transmit their collected sensing results.
We assume that the UAVs in the transmission cycle keep their locations unchanged until their sensing results have been successfully transmitted to the BS.
This is because if the UAV started moving to its next selected task or towards the BS during the transmission cycle, the UAV would have to fly back and sense the target again if the BS ultimately determined the sensing result was invalid.
When the allocated number of subcarriers to UAV $i$ is $k_i$, the amount of data transmitted by UAV $i$ in this cycle is given by
\beq
\label{equ: transmit data in cycle}
D(k_i, \bm x_i) =  k_i\cdot W\cdot \log_2(1+\gamma(\bm x_i))\cdot(t_c - t_e),
\eeq
where $t_e$ denotes the duration of the information exchange at the beginning of the cycle.

\textbf{Remark}: After a UAV has finished the transmission cycles for reporting its sensing results, the BS will inform the UAV of the AoI of the selected task at the beginning of the next cycle.
If the AoI of the selected task is not reduced to one cycle, the UAV knows that its previously transmitted sensing result was invalid.
Then, the UAV will begin another sensing cycle at once and will subsequently transmit the new sensing result.
The sequence of sensing cycle and transmission cycles will be repeated until a valid sensing result is received at the BS.
Moreover, if multiple UAVs are executing the same task, they will consider the task completed when any valid sensing result has been received for the task at the BS.

\section{Problem Formulation}
\label{sec: problem formulation}
In this section, we first formulate the trajectory design problem for the cellular Internet of UAVs as an MDP.
Thereby, the cellular Internet of UAVs represents the environment, and the UAVs are modeled as the agents.
Then, we specify the states, actions, state transitions, and rewards of the MDP.
Finally, we formulate the UAV trajectory design problem for AoI minimization as a distributed optimization problem.

\subsection{Markov Decision Process Formulation}
\begin{figure}[!t] 
	\center{\includegraphics[width=0.55\linewidth]{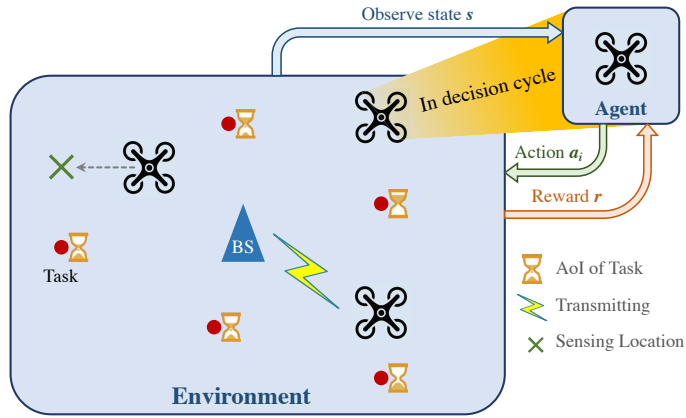}}
	\caption{MDP for distributed trajectory design for the cellular Internet of UAVs.}
	\label{fig: mdp illustration}
	\vspace{-1em}
\end{figure}

The trajectory design problem for the cellular Internet of UAVs can be formulated as an MDP, since the UAVs make sequential decisions regarding their trajectories in cycles, and their trajectory design influences their states as well as {the AoI that the tasks will have in the future}.
As shown in Fig.~\ref{fig: mdp illustration}, the MDP is defined by the considered environment, i.e., the cellular Internet of UAVs and the sensing tasks, and the $M$ agents, i.e., the $M$ UAVs.
 {The basic time step unit for the MDP is a cycle, indicating that the state transitions of the UAV occur at the beginning of every cycle.}
To be specific, we define the tuple $(\mathcal S, \mathcal A, R, \mathcal T)$ for the MDP, which includes the set of states~$\mathcal S$, the set of actions~$\mathcal A$, the reward function~$R$, and the state transition function~$\mathcal T$.
At each cycle, the MDP is in some state $\bm s\in \mathcal S$, and each UAV chooses an action $\bm a\in\mathcal A$ that is available for state $\bm s$.
Then, after one cycle, the process transitions into a new state $\bm s'$ according to the previous state $\bm s$ and the actions taken by the UAVs.
The probability for the state transition between state $\bm s$ and state $\bm s'$ given action $\bm a$ can be deduced from the state transition function $\mathcal T$.
Moreover, after each cycle, the UAVs receive a reward $r=R(\bm s,\bm s')$ from the environment for the state transition that has occurred.
In the following, we specify each element of the formulated MDP, i.e., the states, actions, rewards, and state transitions.

\subsubsection{State}
The state of the environment $\bm s\in\mathcal S$ at the beginning of the $n$-th cycle is defined as 
\beq
\label{equ: state}
\bm s = (
            n, 
            \{\bm x_i\}_{i\in\mathcal M},
            \{D_{r,i}\}_{i\in\mathcal M},
            \{\tau_j\}_{j\in\mathcal N}, 
            \{T_i\}_{i\in\mathcal M}, 
            \{\hat{\bm x}_i\}_{i\in\mathcal M}).
\eeq
Here,
	$n\in[1,N_c]$ is the index of the cycle with $N_c$ being the total number of considered cycles;
	$\bm x_i$ is the location of UAV $i$ at the beginning of the cycle;
	$D_{r,i}$ denotes the amount of sensing data to be transmitted;
	$\tau_{j}$ denotes the AoI of Task $j$;
	$T_i$ denotes the selected task of UAV $i$;
	and $\hat{\bm x}_i$ denotes the corresponding sensing location.
UAVs know the state of the environment from the information exchange with the BS at the beginning of each cycle.
Since the timeline of the adopted protocol is discretized into cycles, $\tau_j$ in~(\ref{equ: state}) is an integer multiple of $t_c$, i.e., the duration of a cycle.
Based on~(\ref{equ: aoi def}), $\tau_j$ can be defined as
\beq
\label{equ: redefind aoi}
\tau_j = n\cdot t_c-\lfloor U_j / t_c\rfloor \cdot t_c,
\eeq
where $\lfloor\cdot\rfloor$ denotes the floor function.

Besides, the type of cycle of UAV $i$ can be expressed as a function of the elements of state $\bm s$, and thus is not modeled as a separate element of the state.
To be specific, based on the proposed protocol, the type of cycle of UAV $i$ at state $\bm s$ is given by
\beq
 {C_i(\bm s)= \begin{cases}
 	C_d, &\text{if } \tau_{T_i} = t_c,\\
 	C_e, &\text{if } \tau_{T_i}\neq t_c\text{ and } \bm x_i \neq \hat{\bm x}_i,\\
 	C_s, &\text{if } \tau_{T_i}\neq t_c,~ \bm x_i = \hat{\bm x}_i, \text{ and } D_{r,i} = 0,\\
 	C_t, &\text{otherwise},\\
 \end{cases}}
\eeq
where $C_d,~C_e,~C_s$, and $C_t$ denote the decision cycle, empty cycle, sensing cycle, and transmission cycle, respectively.

\subsubsection{Action}
We model the action of UAV $i$ as $\bm a_i = (a_{t,i},\bm a_{s,i})$, where $a_{t,i}$ and $\bm a_{s,i}$ denote the selected task and the sensing location of UAV $i$, respectively.
According to the distributed sense-and-send protocol in Section~\ref{sec: protocol}, a UAV makes its decision regarding the selected task and sensing location during the decision cycle and works on the task until a valid sensing result has been successfully received by the BS.
Therefore, the available action set of the UAV in the empty, sensing, and transmission cycles is limited to the given selected task and the corresponding sensing location.
We note that, as the UAVs can be flexibly deployed and can hover in continuous space~\cite{zhang2018joint,Fotouhi2019Survey}, we assume that the sensing locations of the UAVs are continuous variables.

As having multiple UAVs select the same task is inefficient in terms of reducing the average AoI of the tasks, we do not allow the UAVs to select tasks that are currently executed by other UAVs.
Then, the available action set of UAV $i$ at state $\bm s$ can be expressed as follows:
\begin{align}
\label{equ: available action set}
&\mathcal A_i(\bm s) = \\
	&\begin{cases}
		\{(T_i,\hat{\bm x}_i)\},
		~\text{if }C_i \neq C_d,\\
		\{(T,\bm x)| T\in\mathcal N/\{T_{i'}\}_{i'\in\mathcal M, i'\neq i}, \|\bm x-\bm x^t_T\|_2\!\cdot\! \sin\phi\leq r_s\},\\
		\hspace{18em} \text{if } C_i = C_d.
	\end{cases}\nonumber
\end{align}
In the second case of~(\ref{equ: available action set}), the constraint $ \|\bm x-\bm x^t_T\|_2\cdot \sin\phi\leq r_s$ is imposed to ensure that the sensing location is within the maximum sensing angle $\phi$ of the UAV; otherwise, the probability of successful sensing of UAV $i$ will be zero, as indicated in~(\ref{equ: successful task-sensing probability}).

\subsubsection{State Transition Function}
After the $n$-th cycle, the state at the beginning of the next cycle transitions to $\bm s'$.
The state transition function is defined as $\mathcal T:(\bm s, \{\bm a_{i}\}_{i\in \mathcal M})\mapsto {\bm s'}$, where the transition of the elements of $\bm s$ can be expressed as follows:
\begin{subequations}
\label{equ: state transition set}
\begin{align}
&n' = n+1,\\
&\bm x_i' = \bm x_i + \Delta \bm x_i,\\
&D_{r,i}' = \begin{cases}
 	D_s, & \text{if } C_i = C_s,\\
 	\max\{D_{r,i} - D(k_i, \bm x_i),0\}, &\text{if } C_i = C_t,\\
	0,&\text{otherwise},
 \end{cases}\\
&\tau_j' = \begin{cases}
 	t_c,& \text{ {if Task $j$ is executed successfully}},\\
 	\tau_j + t_c,&\text{otherwise},
 \end{cases}\\
&(T_{i}', \hat{\bm x}_i') = \bm a_i,
\end{align}
\end{subequations}
Here,~(\ref{equ: state transition set}a) increases the cycle index.
Eq.~(\ref{equ: state transition set}b) specifies the change of location of the UAV, where the trajectory $\Delta \bm x_i$ can be obtained based on (\ref{equ: next location}).
Eq.~(\ref{equ: state transition set}c) implies that in the sensing cycle, UAV $i$ collects {$D_s$ bits of data}; in the transmission cycle, it transmits $D(w_i, \bm x_i)$ bits of data to the BS; and in the empty and decision cycles, it does not transmit.
Eq.~(\ref{equ: state transition set}d) indicates that if a valid sensing result for Task $j$ is received by the BS, the AoI of the task becomes zero according to the definition of the AoI, otherwise it will increase by the duration of a cycle.
In~(\ref{equ: state transition set}e), $T_{i}'$ and $ \hat{\bm x}_i'$ carry over the selected task and the sensing location of UAV $i$ from the previous state $\bm s$ to the current state $\bm s'$.

Besides, we assume that the BS allocates the same number of subcarriers orthogonally to the UAVs\footnote{ 
The proposed algorithm in Section~\ref{sec: algorithm design} is independent of the subcarrier allocation mechanism and does not rely on the assumption that no interference between the UAVs exists.
Therefore, the proposed algorithm can also be combined with more sophisticated subcarrier allocation mechanisms and can be used regardless of whether interference between UAVs’ exists or not.
}.
Therefore, in~(\ref{equ: state transition set}c), the number of subcarriers allocated to UAV $i$ is given by
\beq
 {k_i = \begin{cases}
 	\lfloor\frac{K}{\sum_{i\in \mathcal M}\mathbb I(C_i=C_t)}\rfloor, &\text{if } C_i= C_t,\\
 	0, &\text{otherwise},
 \end{cases}}
\eeq 
where the indicator function $\mathbb I(\cdot)$ returns $1$ if the condition is true and returns $0$, otherwise.
Since we have assumed that $K>M$ in Section~\ref{sec: system model}, the allocated number of subcarriers to UAV $i$ is always larger than zero, i.e., $k_i>0$.
\subsubsection{Reward}
\begin{figure}[!t] 
	\center{\includegraphics[width=0.7\linewidth]{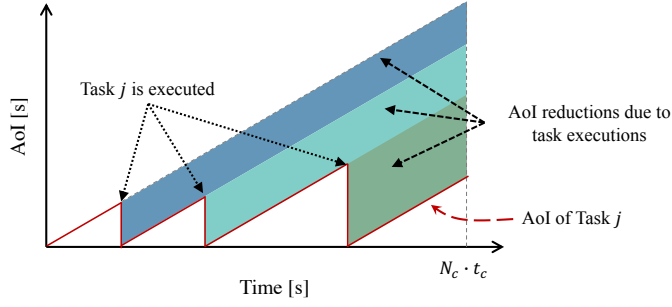}}
\vspace{-1.8em}
	\caption{AoI reduction for Task $j$.}
	\label{fig: aoi reduction}
	\vspace{-1em}
\end{figure}

In general MDPs, the reward is a numerical value obtained by the agent from the environment and quantifies the degree to which the agent's objective has been achieved~\cite{sutton1998reinforcement}.
To define the reward in the cellular Internet of UAVs, we first define the \emph{AoI reduction}.
Denote the AoI of Task $j$ in the $n$-th cycle by $\tau_j(n)$. The accumulated AoI of Task $j$ in the $N_c$ cycles can be calculated as follows:
\begin{align}
\label{equ: def of aoi reduction}
&\sum_{n=1}^{N_c}\! \tau_j(n)= 
\frac{N_c(N_c+1)t_c}{2} \\
&-\! \sum_{n=1}^{N_c}\!\tau_j(n)\!\cdot\!(N_c- n)\!\cdot\! \mathbb I(\text{Task $j$ is executed in the $n$-th cycle}). \nonumber
\end{align}
In~(\ref{equ: def of aoi reduction}), the first term on the right hand side is the maximum accumulated AoI of Task $j$ within the $N_c$ considered cycles which has a constant value, and the second term on the right hand side can be interpreted as the sum of AoI reductions due to the executions of Task $j$, as shown in Fig.~\ref{fig: aoi reduction}.
To be specific, we define the AoI reduction due to the execution of Task $j$ in the $n$-th cycle as $\tau_j(n)\cdot(N_c- n)$.

In the cellular Internet of UAVs, the objective of the UAVs is the minimization of the AoI of the tasks and the UAVs cooperate.
Therefore, we propose that the UAVs share the reward obtained for the AoI reduction in state transition $\bm s' \rightarrow \bm s$, i.e.,
\beq
\label{equ: reward function}
R(\bm s', \bm s) = \sum_{j\in\mathcal N} \tau_{j}'\cdot(N_{c} - n)\cdot \mathbb I(\text{Task $j$ is executed in $\bm s$}),
\eeq
where $\tau_{j}'$ denotes the AoI of Task $j$ in $\bm s'$\footnote{
The proposed reward function~(\ref{equ: reward function}) does not include an explicit punishment for invalid sensing results.
Nevertheless, when a UAV has transmitted an invalid sensing result, it will have to sense the target again and transmit the sensing result until a valid sensing result is received by the BS.
As the AoI of the tasks increases during this repeated sensing and transmission, the rewards for the UAVs for AoI reduction decrease, which serves as the punishment for invalid sensing results.
}.

\subsection{Trajectory Design Policy Optimization}
We define the policy of UAV $i$ in the MDP as a function mapping state $\bm s$ to the action~$\bm a_i$ that UAV $i$ will choose in the state as $\bm \pi_i: \bm s\rightarrow \bm a_i$.
Furthermore, we denote the policy profile of the UAVs by $\bm \pi = (\bm \pi_1,...,\bm \pi_M)$.
The policies of the UAVs indicate how the UAVs design their trajectories when facing different states of the environment.
Therefore, the trajectory design policy determines the performance of the considered cooperative Internet of UAVs and has to be optimized for AoI minimization.

Moreover, for trajectory design, the UAVs should not only consider the current AoI, but also the AoI that the tasks will have in the future.
Therefore, the objective of the proposed trajectory design is to optimize policy $\bm \pi$ for minimization of the \emph{normalized accumulated AoI} in the considered $N_c$ cycles, which can be expressed as 
\begin{align}
\label{equ: normalized AoI}
	&\Psi(\bm \pi) = \frac{
	N\cdot N_c\cdot(N_c+1)\cdot t_c/2 -\sum_{
	\bm s\in \mathcal S^{\bm \pi}_{\mathcal T}(\bm s_0)
	}R(\bm s',\bm s)
	}{N\cdot N_c}, \\
	\label{equ: expectation set}
	&\mathcal S^{\bm \pi}_{\mathcal T}(\bm s_0) \!=\! 	\Big\{\!(\bm s^{(n)}, \bm s^{(n+1)})| \bm s^{(1)} \!=\! \bm s_0, \bm s^{(n+1)}\!=\!\mathcal T(\bm s^{(n)}, \bm\pi(\bm s^{(n)})), \nonumber \\
    &\hspace{16em}n\in[1,N_c)\Big\},
\end{align}
where $\bm s_0$ and $\bm s^{(n)}$ denote a given \emph{initial state} and the state in the $n$-th cycle, respectively.
To be specific, we assume that in state $\bm s_0$, all UAVs start from location $(0,0,h)$ with no stored data for transmission, and the tasks are initialized to have zero AoI\footnote{
The zero initial AoI of each task indicates that the sensing result of each task has been just transmitted to the BSs before the first considered cycle.
Nevertheless, since our proposed problem formulation and algorithms do not require the AoIs of the tasks to take specific values, they can be easily extended to the case where the tasks have different initial AoIs.
}.
Besides, the UAVs have no selected tasks nor corresponding sensing locations.

In~(\ref{equ: normalized AoI}), the sum is taken over all states in $\mathcal S^{\bm \pi}_{\mathcal T}(\bm s_0)$, i.e., all states in the $N_c$ considered cycles.
Specifically, as shown in~(\ref{equ: expectation set}), $\mathcal S^{\bm \pi}_{\mathcal T}(\bm s_0)$ denotes the set of states which are reached from the initial state, $\bm s_0$, within $N_c$ cycles, given the policy of the UAVs, $\bm \pi$, and transition function $\mathcal T$ defined in~(\ref{equ: state transition set}a)-(\ref{equ: state transition set}e).
Besides, in~(\ref{equ: normalized AoI}), the first term in the numerator is the maximum accumulated AoI of the $N$ tasks within $N_c$ cycles, and the second term is the sum of AoI reductions.
Since the maximum accumulated AoI of the tasks is a constant, the maximization of the AoI reduction can be equivalently adopted as the objective for trajectory design.
The resulting optimization problem can be formulated as follows:
\begin{align}
\label{opt: AoI minimization of internet UAVs}
(\text{P}\theoptcnt)\!: \stepcounter{optcnt}
\max_{\bm \pi}~
&\sum_{
	\bm s\in \mathcal S^{\bm \pi}_{\mathcal T}(\bm s_0)
	}R(\bm s), \nonumber\\
 s.t. ~
 &\bm \pi_i(\bm s^{(n)}) \in\mathcal A_i(\bm s^{(n)}), ~
	\forall n\in [1,N_c],~i\in\mathcal M, \nonumber\\
&\mathcal S^{\bm \pi}_{\mathcal T}(\bm s_0) = 	\big\{\bm s^{(n)}|\bm s^{(1)} = \bm s_0, \nonumber \\
&\hspace{3em}\bm s^{(n+1)}=\mathcal T(\bm s^{(n)}, \bm\pi(\bm s^{(n)})), n\in[1,N_c)\big\}. \nonumber
\end{align}
\vspace{-1.5em}

Since the UAVs design their trajectories in a distributed manner, each UAV needs to optimize its own trajectory design policy.
The distributed trajectory optimization problem for UAV $i$ can be formulated as follows:
\begin{align}
(\text{P}\theoptcnt)\!: \stepcounter{optcnt}
\max_{\bm \pi_i}~
&\sum_{\bm s\in \mathcal S^{\bm \pi_i,\bm\pi_{-i}}_{\mathcal T}\!(\bm s_0)}R(\bm s), \nonumber\\
 s.t.~
 &\bm \pi_i(\bm s^{(n)}) \in\mathcal A_i(\bm s^{(n)}), ~\forall n\in [1,N_c],\nonumber\\
 &\mathcal S^{\bm \pi_i,\bm \pi_{-i}}_{\mathcal T}\!(\bm s_0) = \big\{\bm s^{(n)}|\bm s^{(1)} = \bm s_0, \bm s^{(n+1)}\nonumber\\
 &\hspace{1.5em}=\!\mathcal T(\bm s^{(n)},\bm\pi_{i}(\bm s^{(n)}), \bm\pi_{-i}(\bm s^{(n)})), n\!\in\![1,N_c)\big\}\! \nonumber,
\end{align}
where $\bm \pi_{-i}$ denotes the policies of all UAVs except UAV $i$.

Since the sensing and transmission models are not known by the UAVs, the state transition function $\mathcal T$ is unspecified and~(P2) cannot be solved directly.
To overcome this issue, reinforcement learning methods, such as Q-learning~\cite{sutton1998reinforcement}, can be employed, which require no prior information regarding the state transition function.
Hence, we reformulate~(P2) using the concept of the \emph{Q-function}.
Given the policies of the UAVs $\bm \pi$, the Q-function is defined as a mapping from state-action pair $(\bm s,\bm a_i)$ to the total reward of UAV $i$ after taking action $\bm a_i$ in state $\bm s$.
Specifically, the Q-function can be expressed in a recursive manner as follows:
\begin{equation}
\label{equ: q update}
	Q_i^{\bm\pi_i\!,\! \bm\pi_{-i}}(\bm s,\!\bm a_i) \!=\!\begin{cases}
    R(\bm s) \!+\!
	Q_i^{\bm\pi_i\!,\! \bm\pi_{-i}}\!\left(\bm s',\! {\bm \pi}_i(\bm s')\right)|_{
		\bm s'\! = \!\mathcal T(\bm s, \bm a_i, {\bm \pi}_{-i}(\bm s))}
	, \\ \hspace{11.8em} \text{if } n\in[1,N_c),\\
	0,\text{otherwise}.
\end{cases}
\end{equation}
In~(\ref{equ: q update}), the second case is because the total number of considered cycles is $N_c$, and thus the rewards beyond the $N_c$-th cycle are assumed to be $0$.
Based on~(\ref{equ: q update}), for UAV $i$, given the policies of the UAVs, $\bm \pi$, the sum of AoI reduction within the considered $N_c$ cycles is given by $Q_i^{\bm\pi_i, \bm\pi_{-i}}(\bm s_0,\bm \pi_i(\bm s_0))$.
Therefore,~(P2) can be reformulated as
\begin{align}
(\text{P}\theoptcnt):~\stepcounter{optcnt}
\max_{\bm \pi_i}
\quad 
&Q_i^{\bm\pi_i, \bm\pi_{-i}}(\bm s_0,\bm \pi_i(\bm s_0))
\nonumber\\
 s.t. \quad
 &\bm \pi_i(\bm s) \in\mathcal A_i(\bm s),\quad
 	\forall \bm s \in \mathcal S
 . \label{opt: AoI minimization-q version}
\end{align}

Based on~\cite{sutton1998reinforcement}, the optimal policy for problem~(P3) can be obtained as					
\beq
\label{equ: optimal policy}
\bm\pi_i^*({\bm s}) = \argmax_{\bm a_{i}\in\mathcal A(\bm s)} 	~Q^{*, \bm\pi_{-i}}_i({\bm s}, \bm a_i),\quad \forall \bm s\in\mathcal S,
\eeq
where $Q^{*, \bm\pi_{-i}}_i$ denotes the Q-function for UAV $i$ given that the other UAVs adopt policy $\bm \pi_{-i}$ and UAV $i$ adopts its optimal policy $\bm \pi^*_i$.
Therefore, determining the optimal policy in~(\ref{equ: optimal policy}) is equivalent to finding the Q-function $Q^{*, \bm\pi_{-i}}_i({\bm s}, \bm a_i)$.

When the state and action spaces are discrete and small, the optimal Q-function value can be obtained by using classic reinforcement learning methods such as Q-learning~\cite{sutton1998reinforcement}, where a look-up table is maintained for updating the Q-function value based on~(\ref{equ: q update}).
However, in this paper, the state and action spaces are large and involve both discrete and continuous variables.
This means that most state-action pairs will be rarely visited, which results in an unaffordably long time for the Q-function to converge.
Hence, to solve the Q-function efficiently, we adopt deep reinforcement learning, which exploits the generalization capability of deep learning to handle MDPs with large and continuous state and action spaces~\cite{Arulkumaran2017ABrief,goodfellow2016deep}.
In the next section, we introduce the proposed deep reinforcement learning algorithm CA2C for solving~(\ref{equ: optimal policy}).

\section{Compounded-Action Actor-Critic Algorithm}
\label{sec: algorithm design}
In this section, {we propose a deep reinforcement learning algorithm, referred to as CA2C, for the trajectory design problem for the cellular Internet of UAVs.
The CA2C algorithm has the capability to efficiently handle agents who have compounded actions, i.e., actions involving both continuous and discrete variables.}
We first explain the motivation for using the CA2C algorithm, and then we elaborate on the action selection process and the training process.

\subsection{Motivation}
Deep reinforcement learning algorithms have been shown to be suitable for finding the optimal policies for the agents of MDPs with high-dimensional state spaces~\cite{Lillicrap2016Continuous}.
Nevertheless, the existing deep reinforcement learning algorithms cannot be applied directly to the trajectory design problem considered in this paper.
This is because the action of each UAV, {i.e., $\bm a_i = (a_{t,i},\bm a_{s,i})$}, involves both the discrete task selection variable $a_{t,i}\in \mathcal N$ and the continuous sensing location selection variable $\bm a_{s,i}\in\mathbb R^2$, whereas traditional deep reinforcement learning algorithms are designed for agents with purely discrete or purely continuous action spaces.

To tackle this difficulty, we propose the CA2C algorithm, which combines the advantages of the deep deterministic policy gradient~(DDPG) algorithm~\cite{Lillicrap2016Continuous} for handling continuous action spaces and the deep Q-network~(DQN) algorithm~\cite{Volodymyr2015Human} for handling discrete action spaces.

\subsection{Action Selection Process}
To handle the considered MDP whose actions comprise both the discrete task selection and the continuous sensing location selection, we decompose the optimal task selection policy $\bm{ \pi}_i^*$ in~(\ref{equ: optimal policy}) into two parts, i.e., the policy for selecting the optimal task of UAV $i$ and the policy for selecting the optimal sensing location for the task.
Specifically, UAV~$i$ determines the optimal selected task, which maximizes the Q-function value for the current state~$\bm s$, i.e.,
\beq
\label{equ: selected task policy}
a_{t,i}^*= \argmax_{a_{t,i}\in\mathcal N/\{T_{i'}\}_{i'\in\mathcal M, i'\neq i}}Q^{*,\bm \pi_{-i}}_i
	\big( {\bm s}, (a_{t,i}, \bm \nu_i^*({\bm s}, a_{t,i})) \big).
\eeq
Here, policy $\bm \nu_i^*({\bm s}, a_{t,i})$ provides the optimal sensing location given state ${\bm s}$ and selected task $a_{t,i}$.

\begin{figure}[!t] 
	\center{\includegraphics[width=0.7\linewidth]{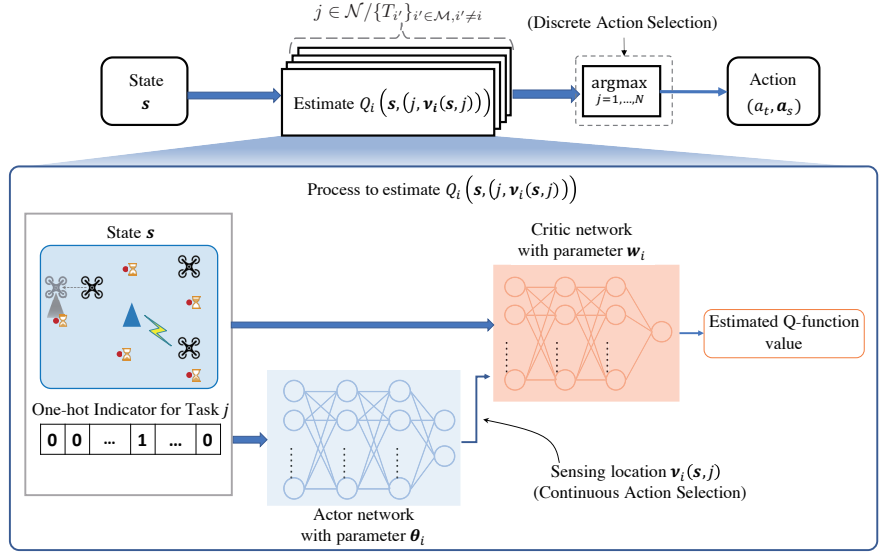}}
	\vspace{-1.2em}
	\caption{Process for the UAV to select the action for state $\bm s$.}
	\label{fig: select action}
	\vspace{-1em}
\end{figure}

Due to the high dimensional state and action spaces, estimating the exact $Q^{*,\bm \pi_{-i}}_i$ and $\bm \nu_i^*$ is time-consuming.
To overcome this problem, we adopt neural networks to approximate $Q^{*,\bm \pi_{-i}}_i$ and $\bm \nu_i^*$, and train the neural networks by combining the training methods for the DQN and DDPG algorithms.
Since in the training process, the continuous action selection~(i.e., the sensing location selection) cannot be trained jointly with the discrete action selection~(i.e., the task selection), we employ two separate neural networks to estimate $Q^{*,\bm \pi_{-i}}_i$ and $\bm \nu_i^*$, respectively, which we refer as the critic network and the actor network, respectively, as illustrated in Fig.~\ref{fig: select action}.

Specifically, Q-function $Q^{*,\bm \pi_{-i}}_i$ is approximated by $Q_i(\bm s, \bm a_i|\bm w_i)$, and the deterministic policy $\bm \nu_i^*$ is approximated by $\bm \nu_i(\bm s, a_{t,i}|\bm \theta_i)$.
Here, $\bm w_i$ and $\bm \theta_i$ denote the parameters of the critic and actor networks, respectively, and comprise the connection weights and the biases of the activation functions in the corresponding neural networks~\cite{Michelucci2018Training}.
When a UAV needs to determine its new selected task, it estimates the Q-function values of the available tasks.
The process for UAV $i$ to estimate the Q-function value of Task $j$ at state $\bm s$ is illustrated in Fig.~\ref{fig: select action}.
UAV $i$ first combines state $\bm s$ and Task $j$ as the input of the sensing location neural network $\bm \nu_i(\bm s, a_{t,i}|\bm \theta_i)$, which derives sensing location $\hat{\bm x}_j$ at its output.
Subsequently, UAV $i$ combines the state, selected task, and sensing location as the input of the critic network, which provides the estimated Q-function value for Task~$j$ as its output.
Comparing all the tasks in terms of the resulting estimated Q-function values, UAV $i$ then determines its optimal selected task based on~(\ref{equ: selected task policy}).

\subsection{Training Process}
The parameters of the critic and actor networks are trained in order to approximate functions $Q^{*,\bm \pi_{-i}}_i$ and $\bm \nu_i^*$.
As in the DDPG and DQN algorithms~\cite{Lillicrap2016Continuous,Volodymyr2015Human}, for the CA2C algorithm each agent, i.e., each UAV, relies on a replay buffer to train its parameters $\bm w_i$ and $\bm \theta_i$.
In the DDPG and DQN algorithms, the replay buffer of an agent stores the agent's experience, i.e., a set of tuples which consist of the state, action, transit state, and reward.
Nevertheless, in the formulated MDP, when the selected task of a UAV has not been completed successfully, the action of the UAV is to continue executing the current selected task.
Therefore, the state transitions between consecutive cycles cannot be used directly as the experience for the training of the UAVs' action selection policies.
To handle this problem, we propose that UAV $i$ records the state transition only during decision cycles, where its previous selected task is completed.
Denote the replay buffer of UAV $i$ by $\mathcal D_i = \{\bm e_i\}$.
The stored experience in the replay buffer is given by $
\bm e_i = (\tilde{\bm s}_i, \tilde{\bm a}_i, \tilde{\bm s}_i', \tilde{r}_i)
$,
where
	$\tilde{\bm s}_i$ denotes the state of the cycle when UAV $i$ is in a certain decision cycle, 
	$\tilde{\bm a}_i$ is the action of UAV $i$ determined in that cycle,
	$\tilde{\bm s}_i'$ denotes the state of the cycle when the selected task is executed,
	and 
	$\tilde{r}_i$ denotes the sum of the rewards for UAV $i$ during the state transition.

For describing the training process, we refer to the considered $N_c$ cycles as an \emph{episode}.
Within an episode, a UAV stores experience $\bm e_i$ in its replay buffer after it has executed its selected task.
At the end of an episode, each UAV $i$ trains its actor and critic networks based on a batch of $N_{b}$ experiences, which are sampled randomly from $\mathcal D_i$ and denoted by $\mathcal B_i$.
The training process of the actor and critic networks is described in the following.

Based on~\cite{Lillicrap2016Continuous}, parameter $\bm\theta_i$ in the actor network is updated by taking steps in the direction of the gradient of the performance evaluation function $J_i(\bm \theta_{i})$,
	i.e.,
\beq
\label{equ: theta update}
\bm \theta_i = \bm \theta_i + \alpha \nabla_{\bm \theta_{i}} J_i(\bm \theta_{i}),
\eeq
where $\alpha$ denotes the learning rate.
Here, $J_i(\bm \theta_{i})$ can be expressed as
\beq
J_i(\bm\theta_i) = 
	\frac{1}{N_b}\sum_{\bm e_i \in \mathcal B_i}
		Q_i(
				\tilde{\bm s}_i, 
				(\tilde{a}_{t,i}, \bm\nu_i(\tilde{\bm s}_i, \tilde{a}_{t,i}))|\bm \theta_i)
			),
\eeq
and the gradient $\nabla_{\bm \theta_{i}} J_i(\bm \theta_{i})$ can be calculated as
\begin{align}\label{equ: RL policy gradient}
\nabla_{\bm \theta_{i}} J_i(\bm \theta_{i}) = 
	&\frac{1}{N_b}\sum_{\bm e_i \in \mathcal B_i}
	\nabla_{\bm \theta_{i}} \bm\nu_i(\tilde{\bm s}_i, \tilde{a}_{t,i}|\bm\theta_{i})  \\
    &\times
	\nabla_{\bm a_{s,i}}Q_i
		\big(
			\tilde{\bm s}_i, (\tilde{a}_{t,i}, \bm a_{s,i}|\bm w_i)
		\big)\big|_{\bm a_{s,i} = \bm \nu_i(\tilde{\bm s}_i, \tilde{a}_{t,i}|\bm\theta_{i}) }, \nonumber
\end{align}
where the sum is taken over sample batch $\mathcal B_i$ of the experiences in $\mathcal D_i$.

\begin{figure}[!t] 
	\center{\includegraphics[width=0.7\linewidth]{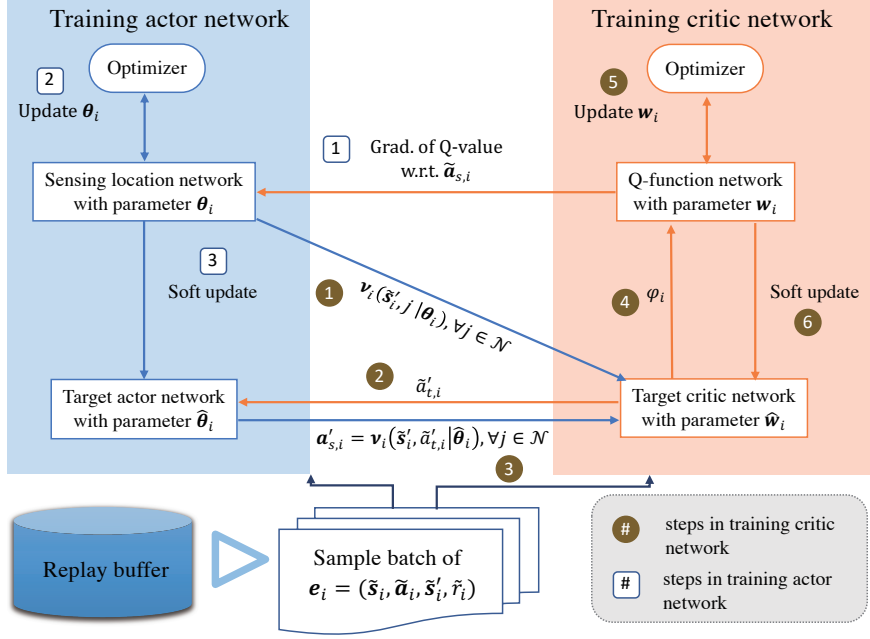}}
	\caption{Training process of the actor network and the critic network.}
	\label{fig: training process}
	\vspace{-0.5em}
\end{figure}

As for the training process of the critic network, the Q-function parameter $\bm w_{i}$ is updated in a similar manner as $\bm \theta_i$ in (\ref{equ: theta update}). The corresponding loss function can be expressed as follows:
\begin{align}
\label{equ: q-function loss}
\mathcal L_i(\bm w_i) &= 
\frac{1}{N_b}\sum_{\bm e_i \in \mathcal B_i}
	\left(Q_i(\tilde{\bm s}_i,\tilde{\bm a}_i|\bm w_i)-\varphi_i(\bm e_i)\right)^2,\\
\label{equ: y}
\varphi_i(\bm e_i) &= \tilde{r}_i +  Q_i
	\big(
		\tilde{\bm s}_i', (a'_{t,i}(\bm e_i),\bm\nu_i(\tilde{\bm s}_i', a'_{t,i}|\hat{\bm \theta}_i)) |\hat{\bm w}_i
	\big), \\
\label{equ: double dqn}
	a'_{t,i}(\bm e_i) &= \argmax_{j\in\mathcal N}Q_i
	\big(
		\tilde{\bm s}_i', (j, \bm \nu_i(\tilde{\bm s}_i',j|\bm \theta_i) )|{\bm w}_i
	\big).
\end{align}
Here, $\hat{\bm w}_i$ and $\hat{\bm \theta}_i$ denote the parameters of the target critic and target actor networks, respectively, which are delayed replicates of the critic and actor networks.
To be specific, in~(\ref{equ: double dqn}) the action for the next state is chosen according to the trained networks, i.e., the networks with parameters $\bm w_i$ and $\bm\theta_i$, while the estimated Q-value for the state-action pair in the next state is estimated using target networks, i.e., the networks with parameters $\hat{\bm w}_i$ and $\hat{\bm \theta}_i$.
This approach, which is referred to as the double Q-learning method, was proposed in~\cite{van2016deep1} to fix the tendency of traditional Q-learning to overestimate the Q-values and improve the training efficiency.
Besides, to calculate $\nabla_{\bm \theta_{i}} J_i(\bm \theta_{i})$ and $\nabla_{\bm w_i}\mathcal L_i(\bm w_i)$, the back-propagation algorithm~\cite{Michelucci2018Training} can be employed.

The overall training process is illustrated in Fig.~\ref{fig: training process}, and the training processes of the actor and critic networks are summarized in Algorithm~\ref{alg: training q fun. net.} and Algorithm~\ref{alg: training sens. loc. net.}, respectively.
The proposed CA2C algorithm is summarized in Algorithm~\ref{alg: cac}.

\setlength{\textfloatsep}{3pt}
\begin{algorithm}[!t]  \label{alg: training q fun. net.}
\small
  \caption{Training algorithm for critic network}
  \begin{algorithmic}[1]  
  \Require{
  Experience sample $\tilde{\bm e}_i=(\tilde{\bm s}_i, \tilde{\bm a}_i, \tilde{r}_i, \tilde{\bm s}'_i)$;
  parameter $\bm w_i$;
  target parameter $\hat{\bm w}_i$;
  soft update parameter $\beta$.
  }
  \Ensure{
  Updated $\bm w_i$ and $\hat{\bm w}_i$.
  }\\
	Obtain the sensing locations for Task $j$, $\forall j\in\mathcal N$, for state $\tilde{\bm s}_i'$.\\
	Based on the Q-function value obtained by the target critic network, determine selected task $a_{t,i}'$ for state $\tilde{\bm s}_i'$.\\
	The target actor network calculates the sensing location for selected task $a_{t,i}'$ for state $\tilde{\bm s}_i'$.\\
	Based on~(\ref{equ: y}), obtain $\varphi_i$ by adding $\tilde{r}_i$ and the output Q-function value of the state-action pair, i.e., $ Q_i( a'_{t,i},\bm\nu_i(\tilde{\bm s}_i', a'_{t,i}|\hat{\bm \theta}_i) )$, obtained from the target critic network.\\
	Update $\bm w_i$ by using the optimizer in the critic network as $\bm w_i = \bm w_i - \alpha\nabla_{\bm w_i}\mathcal L_i(\bm w_i)$.\\
	Update parameter $\hat{\bm w}_i$ in the target critic network via soft update $\hat{\bm w}_i = (1-\beta)\hat{\bm w}_i + \beta {\bm w}_i$.
\end{algorithmic}  
\end{algorithm}

\begin{algorithm}[!t]  \label{alg: training sens. loc. net.}
\small
  \caption{Training algorithm for actor network}
    \begin{algorithmic}[1]  
  \Require{
  Experience sample $\tilde{\bm e}_i=(\tilde{\bm s}_i, \tilde{\bm a}_i, \tilde{r}_i, \tilde{\bm s}'_i)$;
  parameter $\bm \theta_i$;
  target parameter $\hat{\bm \theta}_i$;
  soft update parameter $\beta$.
  }
  \Ensure{
  Updated $\bm \theta_i$ and $\hat{\bm \theta}_i$.
  }\\
	Calculate the gradients of the Q-function value with regard to the sensing locations for all the sampled experiences, i.e., $\nabla_{\bm a_{s,i}} Q_i(\tilde{\bm s}_i,(\tilde{a}_{t,i},{\bm a}_{s,i})|\bm w_i)|_{\bm a_{s,i} = \tilde{\bm a}_{s,i}}$, $\forall \bm e_i \in\mathcal B_i$\\
	Update $\bm\theta_i$ by using the optimizer in the actor network as $\bm\theta_i = \bm\theta_i + \alpha\nabla_{\bm \theta_{i}} J_i(\bm \theta_{i})$.\\
	Update parameter $\hat{\bm \theta}_i$ in the target actor network via soft update $\hat{\bm \theta}_i = (1-\beta)\hat{\bm \theta}_i +\beta {\bm \theta}_i$.
\end{algorithmic}  
\end{algorithm}

\begin{algorithm}[!t]  \label{alg: cac}
\small
	\label{alg: CA2C}
  \caption{CA2C Algorithm}
  \begin{algorithmic}  
  \Require{
	Exploration parameter $\epsilon$;
	initial parameters $\bm w_i$, $\bm \theta_i$, $\hat{\bm w}_i$, $\hat{\bm \theta}_i$, ~$\forall i\in\mathcal M$;
	replay buffer $\mathcal D_i = \emptyset$,~$\forall i \in\mathcal M$;
	maximum number of training episodes $N_{ep}$;
	number of considered cycles $N_c$.
  }
  \Ensure{
  Trained parameters $\bm w_i$, $\bm \theta_i$, ~$\forall i\in\mathcal M$.
  }
  
  Initialize $\bm w_i$, $\bm \theta_i$, and $\mathcal D_i$ for each UAV $i$.
   
   \For{$n_{ep} = 1$ to $N_{ep}$}
   {   
	   \For{$n = 1$ to $N_c$}
	   {
	 		\For{$i\in\mathcal M$}
	 		{
		 		Extract the state $\bm s$ from the BS beacon.	 			
	 			
	 			\If{$\tau_{T_i}=0$}
	 			{
	 				
	 				With probability $\epsilon$, choose a random task as the selected task, and otherwise, choose  
	 				
	 				~~selected task $a_{t,i}$ based on (\ref{equ: selected task policy}).
	 				
	 				Determine the sensing location for the selected task as $\bm a_{s,i} = \bm\nu_i({\bm s},a_{t,i}|\bm\theta_i)$.

	 			}
	 			\Else
	 			{
	 				Continue the selected task by performing action $\bm a_i = (T_i, \hat{\bm x}_i)$.
	 			}	 			
	 		}
	 		
	 		Transition to new state ${\bm s}'$ based on $\mathcal T$, and get reward $R(\bm s)$.
	 		
	 		\For{$i\in\mathcal M$}
	 		{
	 			Accumulate reward $\tilde{r}_i = \tilde{r}_i + R(\bm s)$.
	 				 			
	 			\If{$\tau_{T_i'}=0$}
	 				{
	 					Store experience $\bm e_i$ in buffer $\mathcal D_i$.					
						
						Set $\tilde{r}_i = 0$.
	 				}
	 		}

	    }
	    \For{$i\in\mathcal M$}
	    {
	    	Train the critic network of UAV $i$ by invoking Algorithm~\ref{alg: training q fun. net.}.
	    	
			Train the actor network of UAV $i$ by invoking Algorithm~\ref{alg: training sens. loc. net.}.
	   }
}
\end{algorithmic}
\end{algorithm}
   
\setlength{\textfloatsep}{18pt}

\section{Algorithm Analysis}
In this section, we analyze the computational complexity and the convergence of the proposed CA2C algorithm.
\subsection{Computational Complexity}
Since the CA2C algorithm consists of two main parts, i.e., the action selection and the training process, we analyze their respective computational complexity in the following.
\subsubsection{Complexity of Action Selection}
In the proposed CA2C algorithm, the computationally most expensive part is the determination of the actions of the UAVs, i.e., the selected tasks and the corresponding sensing locations, during the decision cycle.
For each UAV $i$, the computational complexity to determine its action is given in Theorem~\ref{theo: action determine complexity}.

\begin{theorem} \label{theo: action determine complexity}
\emph{(Computational Complexity of Action Selection)}
For UAV $i$, the computational complexity to determine its action in the decision cycle by~(\ref{equ: selected task policy}) is $\mathcal O(N^2\cdot M)$.
\end{theorem}

\begin{IEEEproof}
For a fully connected neural network with fixed numbers of hidden layers and fixed numbers of neurons in the hidden layers, the computational complexity to calculate the output given an input is proportional to the sum of the sizes of input and output~\cite{sipper1993serial}.
Based on~(\ref{equ: state}) and~(\ref{equ: selected task policy}), the sizes of the inputs of the critic and actor networks are $1+5M+N+M\cdot N$ and $3+5M+2N+M \cdot N$, respectively, where the term $M\cdot N$ is caused by the one-hot vector encoding of the $M$ selected tasks.
Therefore, the computational complexity to select the sensing location and to estimate the Q-function value for a state-action pair is $\mathcal O(N\cdot M)$.
Since UAV~$i$ needs to estimate the Q-function values of all $N$ tasks, the computational complexity for the action selection is $\mathcal O(N^2\cdot M)$.	
\end{IEEEproof}

\subsubsection{ {Complexity of Training Process}}
 {For the proposed CA2C algorithm, the computational complexity of the training process is given in Theorem~\ref{theo: training complexity}.
\begin{theorem}\label{theo: training complexity}
\emph{(Computational Complexity of Training Process)}
Given the size of the training batch $N_{b}$, the computational complexity for a UAV to train its actor and critic networks with the CA2C algorithm is $\mathcal O(N_{b}\cdot N^2\cdot M)$. \end{theorem}}

\begin{IEEEproof}
As suggested in~(\ref{equ: double dqn}), before the actual training, the UAV needs to calculate and compare the Q-function values of the $N$ tasks in the transitioned state for each sampled experience. 
Based on Theorem~\ref{theo: action determine complexity}, the computational complexity for this step is $\mathcal O(N_{b}\cdot N^2\cdot M)$.
Besides, for a fully connected neural network with fixed numbers of hidden layers and neurons, the computational complexity of the back-propagation algorithm is proportional to the product of the input size and the output size~\cite{sipper1993serial}.
The sizes of the inputs of the critic and actor networks are $1+5M+N+M\cdot N$ and $3+5M+2N+M \cdot N$, respectively. 
The sizes of the outputs of the critic and actor networks are $1$ and $2$, respectively.
Therefore, the computational complexity of the back-propagation algorithm is $\mathcal O(N_{b}\cdot N\cdot M)$.
In summary, the computational complexity of the training process is $\mathcal O(N_{b}\cdot N^2\cdot M)$.
\end{IEEEproof}

For comparison, we also analyze the computational complexities of the action selection and training processes of the DQN~\cite{Volodymyr2015Human} and DDPG~\cite{lowe2017multi} algorithms.
In the DQN algorithm, each UAV in the decision cycle employs a critic network to estimate the Q-function values corresponding to the $N$ tasks, and selects the task with the maximum Q-value as its next selected task.
As the DQN cannot handle continuous actions, i.e., the sensing location selection, we assume that the UAV selects the location right above its selected task at altitude $h$ as its sensing location, and thus no actor network is needed for the DQN algorithm.
The output size of the critic network in the DQN algorithm is $N$.
Besides, based on~(\ref{equ: state}), the input size of the critic network is $1+5M+N+M\cdot N$.
Then, based on~\cite{sipper1993serial}, the computational complexity of the task selection in the DQN algorithm is $\mathcal O(N^2\cdot M)$.
Moreover, similar to the analysis in Theorem~\ref{theo: training complexity}, the computational complexity of the training process can be shown to be $\mathcal O(N_b\cdot N^2\cdot M)$.

In the DDPG algorithm, each UAV in the decision cycle adopts an actor network to jointly determine its next selected task and the corresponding sensing location, which leads to an input size of to $1+5M+N+M\cdot N$ and an output size of $N+2$.
Besides, in the training process, each UAV employs a critic network to estimate the Q-function values of the state-action pairs, which has an input size of $1+5M+N+M\cdot N$ and an output size of $1$.
Then, based on~\cite{sipper1993serial} and the analysis in Theorem~\ref{theo: training complexity}, the computational complexities of the action selection and training processes of the DDPG algorithm are $\mathcal O(N^2\cdot M)$ and $\mathcal O(N_b\cdot N\cdot M)$, respectively.

In summary, compared to the DQN and DDPG algorithms, the proposed CA2C algorithm has a similar computational complexity.
Besides, as shown in Section~\ref{sec: simulation results}, the proposed CA2C algorithm outperforms the DDPG and DQN algorithms, since it is specifically tailored for the cellular Internet of UAVs where the UAVs have both continuous and discrete action spaces.

\subsection{Convergence}
From (\ref{equ: selected task policy}), it can be observed that the proposed algorithm is essentially an extended Q-learning algorithm. The convergence of the Q-learning algorithm is analyzed in Theorem~\ref{theo: convergence of Q}.

\begin{theorem} \label{theo: convergence of Q}
\emph{(Convergence of Q-learning Algorithm \cite{watkins1992q})}
Consider an MDP which consists of state $\bm s\in\mathcal S$, action $\bm a \in\mathcal A$, and reward function $R(\bm s,\bm a)$.
Denote the learning rate in the $t$-th training step by $\alpha^{(t)}$. The Q-learning algorithm given by 
\begin{equation}
Q(\bm s,\bm a) \! =\!  Q(\bm s,\bm a)
+ \alpha^{(t)}\! \left( R(\bm s,\bm a) +  \max_{\bm a'}Q(\bm s',\bm a')  \!-\! Q(\bm s,\bm a) \right)\nonumber
\end{equation}
will converge to the optimal Q-function which maximizes the value for each state-action pair as $t\rightarrow\infty$ with probability $1$, if $\sum_{t=0}^{\infty}\alpha^{(t)} = \infty$,  $\sum_{t=0}^{\infty} \left(\alpha^{(t)}\right)^2<\infty$, and $|R(\bm s,\bm a)|$ is bounded.
\end{theorem}
\begin{IEEEproof}
Please refer to \cite{watkins1992q}.	
\end{IEEEproof}

We note that, in general, the inverse time decaying learning rate is an effective technique to train neural networks~\cite{You2019How}.
In the first training epochs, the large learning rate accelerates training and prevents the network from being trapped in a bad local optimum near the initial point.
As the number of training epochs increases, the shrinking learning rate helps the network converge to a local optimum and avoid oscillation.

Nevertheless, Theorem~\ref{theo: convergence of Q} only guarantees the convergence of the Q-learning algorithm in the single-agent case, i.e., there is only one agent who adopts the Q-learning algorithm in the environment.
The convergence of the Q-learning algorithm for the multi-agent case is guaranteed only in special cases, such as iterated dominance solvable games and team games \cite{bowling2003multiagent}.
As multiple UAVs adopt the proposed CA2C algorithm and learn simultaneously in the cooperative cellular Internet of UAVs, the convergence is hard to prove.
We analyze the convergence of the proposed CA2C algorithm through simulations in Section~VII.

\section{Simulation Results}
\label{sec: simulation results}

\begin{table}[!t]
\caption{Simulation Parameters}
\centering
\scriptsize
\begin{tabular}{| c |c|}

\Xhline{1.pt}
\textbf{Parameter}	&	\textbf{Value}\\
\hline
\hline
Number of Tasks ($N$) & 10 \\
\hline
 {Maximum Speed of UAVs ($v_{\max}$)} &  {15 m/s}\\
\hline
 {Altitude of UAVs ($h$)} &  {200 m} \\
\hline
 Height of BS ($H_0$) & 25 m\\
\hline
UAV Transmit Power ($P$) & 23 dBm \\
\hline
 {Noise Power ($N_0$)} &  {-96 dBm}\\
\hline
Carrier Frequency ($f_c$) & 2 GHz \\
\hline
Bandwidth of Subcarrier ($W$) & $12.5$~kHz \\
\hline
Duration of a Cycle  ($t_c$) & 100 ms \\
\hline
Duration of Information Exchange ($t_e$) & 20 ms\\
\hline
Number of Subcarriers ($K$) & 80 \\
\hline
 Sensing Parameter ($\lambda$) & 0.01\\
\hline
Maximum Sensing Angle ($\phi$) & $30^\circ$ \\
\hline
Data Size of Sensing Result ($D_s$) & $10^6$ Bytes \\
\hline
Exploration Ratio ($\epsilon$) & 0.1 \\
\hline
Coverage of BS ($R_c$) & 500 m \\
\hline
 {Number of Cycles per Training Episode ($N_{c}$)} &  {$8\times 10^3$}\\
\hline
Sampling Batch Size ($N_{b}$) & $256$\\ 
\hline
Soft Update Parameter ($\beta$) & $0.01$ \\
\hline
Initial Learning Rate ($\alpha_0$) & 0.1\\
\hline
Decaying Rate for Learning Rate ($\eta$) & $10^{-3}$\\
\Xhline{1.pt}
\end{tabular}
\label{T1}
\vspace{-1em}
\end{table}

In this section, we provide simulation results {in order to verify the effectiveness of the proposed CA2C algorithm.
Besides, we also evaluate the impact of UAV cooperation, the number of UAVs, the number of subcarriers, and the flying altitude on the performance of the cellular Internet of UAVs in terms of AoI minimization.}
The employed actor and critic neural networks are three-layer fully-connected networks with $512$ neurons in each layer, and rectified linear units (ReLUs) are used as activation functions.
Moreover, the Adam optimizer~\cite{kingma2014adam} is adopted for the training of the actor and critic networks.
The simulation parameters are summarized in Table~\ref{T1}.

\begin{figure}[t]
	\center{\includegraphics[width=2.7in]{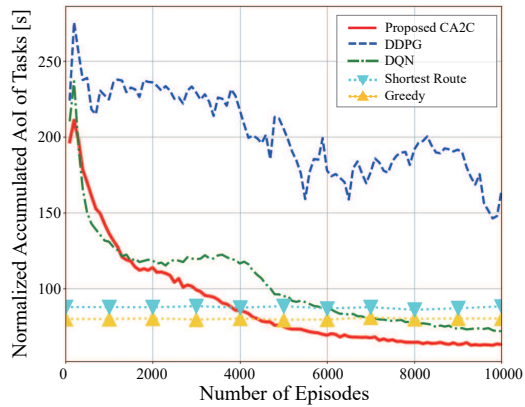}}
	\vspace{-.5em}
	\caption{ {Normalized accumulated AoI of tasks versus the number of episodes for different algorithms. For clarity of presentation, each point of the curves represents the mean normalized accumulated AoIs over the previous $100$ episodes.}}
	\label{fig: training curves}
	\vspace{-1em}
\end{figure}

In the simulation, we compare the proposed algorithm with four baseline algorithms.
We first compare the learning performance of the proposed CA2C, the DDPG algorithm in~\cite{lowe2017multi}, and the DQN algorithm~in\cite{Volodymyr2015Human}.
Furthermore, we compare the proposed algorithm with two conventional, non-learning based algorithms, namely the \emph{greedy algorithm} and the \emph{shortest route algorithm}.
These four benchmark algorithms are briefly described below:
\begin{itemize}[leftmargin=*]
\item \textbf{DDPG}: The action of each UAV is represented as a continuous $(N+2)$-dimensional vector, where the first $N$ dimensions give the respective probabilities for the $N$ tasks to be selected.
\item \textbf{DQN}: The UAVs choose the location right above the selected task as the sensing location.
\item \textbf{Greedy}: In the decision cycle, the UAVs select the task with the highest AoI and move to the nearest location within the sensing range of the tasks.
\item \textbf{Shortest Route}: Each UAV starts from the sensing location of a random task and circles around this location on the shortest route that travels sequentially through the sensing ranges of the other $N-1$ tasks in order to sense the targets and transmit the sensing results. The shortest route is found by the dynamic programming algorithm~\cite{bellman1962dynamic}, which optimizes the order for task selection, and the gradient descent algorithm, which optimizes the sensing locations of the tasks.
\end{itemize}

Fig.~\ref{fig: training curves} shows the normalized accumulated AoI of the tasks, as defined in~(\ref{equ: normalized AoI}), versus the number of training episodes for different algorithms.
Since the connection weights and biases for the neural networks are initialized randomly for all three learning-based algorithms, the normalized accumulated AoIs obtained by these algorithms are similar for the first few training episodes.
It can be observed that the proposed CA2C algorithm has the highest training speed and converges to the lowest normalized accumulated AoI value among the reinforcement learning algorithms.
Compared to the DQN algorithm, the CA2C algorithm has a higher training speed for the following reason.
In the proposed CA2C algorithm, the Q-function values for the tasks change more quickly compared to the DQN algorithm due to the update of the sensing locations.
This increases the probability that a UAV in a certain state tries out different tasks rather than keeping selecting the current best task, which prevents the UAVs from getting trapped in a locally optimal task selection and helps the UAVs explore the optimal task selection policies.
Besides, since the sensing locations are selected for AoI minimization instead of being fixed, the CA2C algorithm also converges to a lower normalized accumulated AoI compared to the DQN algorithm.

We can also observe in Fig.~\ref{fig: training curves} that the DDPG algorithm has the slowest training speed and does not converge within $10,000$ episodes, which can be explained as follows.
In the DDPG algorithm, the UAV determines the selected task based on a probability vector, which is updated during each training process with a small learning rate~($10^{-4} - 10^{-2}$) to stabilize the training~\cite{Mnih2016Asynchronous}.
Due to the small learning rate, the change in the probability vector during one update is small.
Hence, a large number of training episodes is needed before a significant change of the probability vector can be observed.
Therefore, the training speed of the DDPG algorithm is slow.

Moreover, compared to the two non-learning based algorithms, the CA2C algorithm results in a lower normalized accumulated AoI of tasks after $6000$ episodes of training.
This is because with the conventional algorithms, the UAVs determine their actions based on partial information of the environment.
Specifically, in the greedy algorithm, the UAVs only consider the current AoIs of the tasks, and in the shortest route algorithm, the UAVs only consider the distance between their sensing locations.

\begin{figure}[!t]
\center{\includegraphics[width=2.7in]{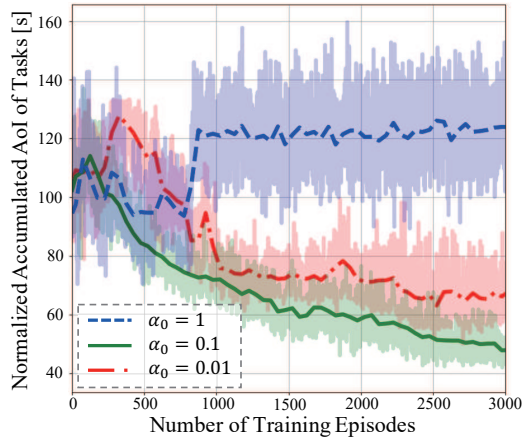}}
	\vspace{-.5em}
	\caption{ {Normalized accumulated AoI of tasks for the proposed CA2C algorithm versus the number of training episodes for different values of initial learning rate $\alpha_0$ and $M=2$. For each case, the points on the darker line represent the mean normalized accumulated AoIs over the previous $50$ training episodes, and the shaded lighter region represents the corresponding variance.}}
	\label{fig: lr compare}
	\vspace{-1em}
\end{figure}

Fig.~\ref{fig: lr compare} shows the normalized accumulated AoI of the tasks for the proposed CA2C algorithm versus the number of training episodes, for different values of initial learning rate $\alpha_0$.
As can be observed, when the initial learning rate is too large, i.e., $\alpha_0 = 1$, the training performance is poor, and the normalized accumulated AoI oscillates at high values.
This is due to fact that large learning rates lead to rapid changes in the parameters of the critic and actor networks, which makes the training process unstable.
Besides, when $\alpha_0$ is too small, i.e., $\alpha_0=0.01$, the training speed is lower than when $\alpha_0=0.1$.
Because the smaller learning rate results in smaller changes of the critic and actor networks' parameters in each episode and leads to a lower training speed.

\begin{figure}[!t]
	\center{\includegraphics[width=2.7in]{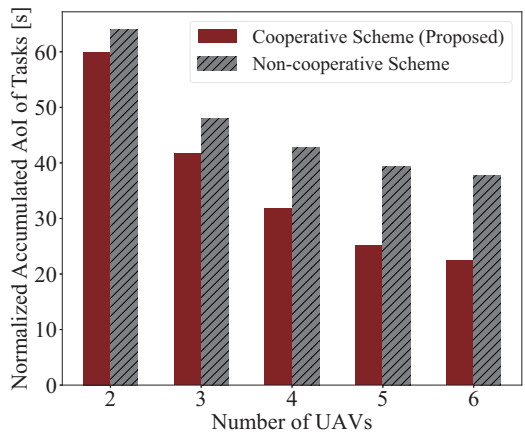}}
		\vspace{-.5em}
	\caption{Normalized accumulated AoI versus the number of UAVs for the cooperative and non-cooperative scheme.}
	\label{fig: influence of number of uavs}
	\vspace{-1em}
\end{figure}

In Fig.~\ref{fig: influence of number of uavs}, we investigate the cooperation gain by comparing the performances of the non-cooperative Internet of UAVs and the cooperative Internet of UAVs, which are referred to as ``\emph{non-cooperative scheme}" and ``\emph{cooperative scheme}" in the figure, respectively.
To be specific, in the non-cooperative scheme, the UAVs do not take into account the trajectories of the other UAVs, and the reward for each UAV is its own AoI reduction rather than the shared reward in~(\ref{equ: reward function}).
For a fair comparison, the UAVs for both the non-cooperative and cooperative schemes employ the proposed CA2C algorithm for trajectory design.

Fig.~\ref{fig: influence of number of uavs} shows the normalized accumulated AoIs of the tasks versus the number of UAVs for the cooperative and non-cooperative schemes, respectively.
As the number of UAVs increases, the normalized accumulated AoI decreases for both schemes.
This is because a larger number of UAVs enables the parallel execution of more tasks, which reduces the normalized accumulated AoI.
Besides, compared to the non-cooperative scheme, the proposed cooperative scheme achieves a lower normalized accumulated AoI of the tasks.
Moreover, the cooperation gain in terms of the AoI increases with the number of UAVs, which can be explained as follows.
In the cooperative scheme, the UAVs cooperate for AoI minimization of all tasks.
However, in the uncooperative scheme, multiple UAVs tend to select the task which provides the highest AoI reduction.
Therefore, the non-cooperative scheme cannot benefit from a large number of UAVs to the same extent as the cooperative scheme.
As a result, the cooperation gain in terms of the AoI increases with the number of UAVs.

\begin{figure*}[!t]
\centering
	\center{\includegraphics[width=1\linewidth]{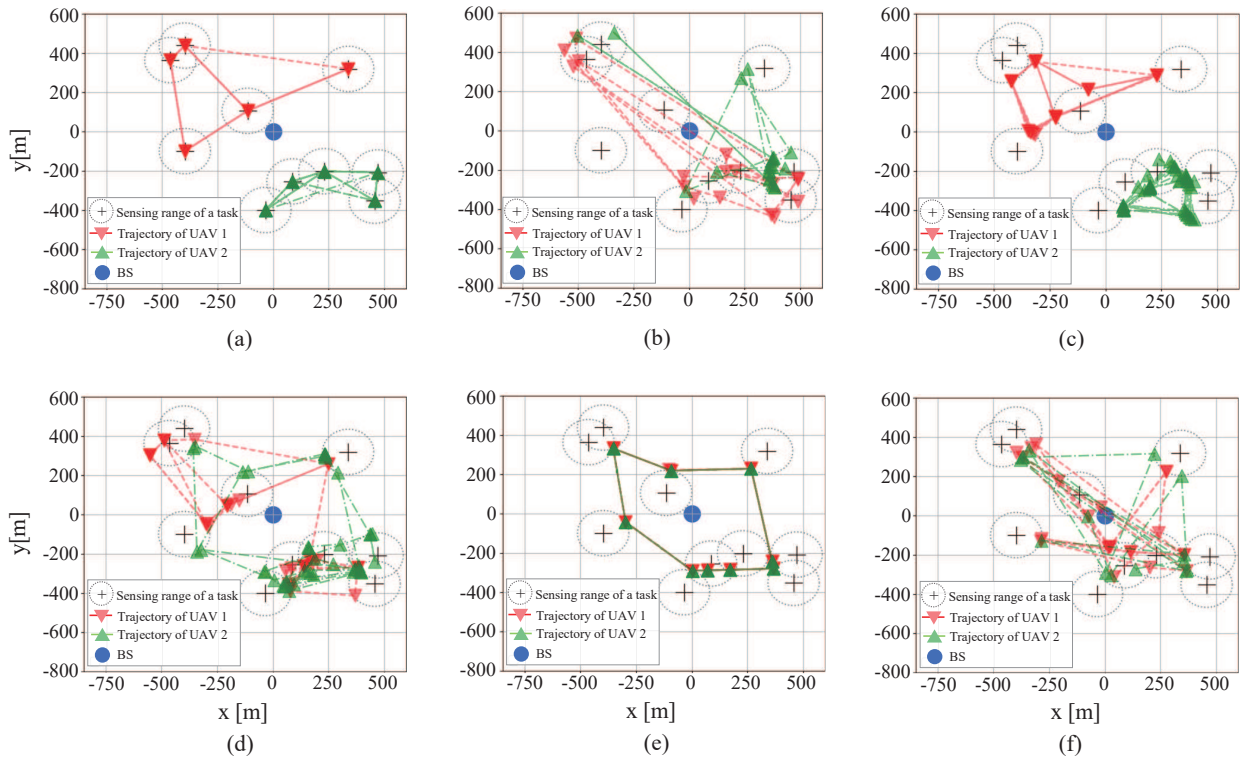}}
		\vspace{-1em}
	\caption{ {Trajectories of two UAVs during one episode. In~(a),~(b), and~(c)~the UAVs are cooperative and trained with the DQN, DDPG, and CA2C algorithms, respectively. In~(d), the UAVs are non-cooperative and trained with the CA2C algorithm. In~(e) and~(f), the two UAVs adopt the shortest route algorithm and the greedy algorithm, respectively.}}
		\label{fig: cruise trajectory}
\vspace{-1em}
\end{figure*}

Fig.~\ref{fig: cruise trajectory} shows the trajectories of the UAVs trained with different algorithms during one episode, where $M=2$. 
In Figs.~\ref{fig: cruise trajectory}~(a),~(b), and~(c), the UAVs are cooperative and trained with the DDPG, the DQN, and the proposed CA2C algorithms, respectively.
In Fig.~\ref{fig: cruise trajectory}~(d), the UAVs are non-cooperative and trained with the CA2C algorithm.
 {
Besides, in Figs.~\ref{fig: cruise trajectory}~(e) and~(f), the UAVs employ the non-learning algortihms, i.e., the greedy algorithm and the shortest-route algorithm, respectively.
}

Fig.~\ref{fig: cruise trajectory}~(a) shows that the two cooperative UAVs trained with the DQN algorithm have learned to divide the tasks into two non-overlapping sets, and each UAV executes the tasks in one of the sets, which results in a low normalized accumulated AoI.
However, since sensing location selection is not performed in the DQN algorithm, the sensing location for a given target is right above the target.
Therefore, the UAVs need to fly long distances between two sensing locations.

Fig.~\ref{fig: cruise trajectory}~(b) shows that the DDPG algorithm also trains the UAVs to select the sensing locations. 
Nevertheless, the task selection of the UAVs is inefficient.
Unlike for the DQN algorithm, the tasks have not been partitioned into separate sets.
Besides, some tasks are left unattended and are not sensed during the episode, which results in a high AoI for these tasks.
This inefficiency of the DDPG algorithm is caused by modeling the task selection by a task-selection probability vector.
This probability vector is difficult to train since the influence of small changes in the task selection probabilities takes a long time to be fully captured.

Fig.~\ref{fig: cruise trajectory}~(c) shows that the UAVs trained with the proposed CA2C algorithm have learned to divide the tasks into two non-overlapping set and select sensing locations which are close to each other.
Therefore, compared to the DQN algorithm, the proposed CA2C algorithm results in shorter flight distances for the UAVs and lower normalized accumulated AoIs of the tasks.

In Fig.~\ref{fig: cruise trajectory}~(d), the sets of tasks executed by the two non-cooperative UAVs are overlapping because the UAVs are competing with each other to maximize their own AoI reduction.
Therefore, each UAV may attempt to execute the task which results in the largest AoI reduction, even if the other UAV is closer to that task.
This inefficiency results in a higher normalized accumulated AoI for the non-cooperative scheme compared to the cooperative scheme.

In Fig.~\ref{fig: cruise trajectory}~(e), it can be observed that, with the shortest route algorithm, the two UAVs circle the same shortest route to execute the $N$ tasks sequentially. 
Besides, as shown in Fig.~\ref{fig: cruise trajectory}~(f), the greedy algorithm results in overlapping UAV trajectories.
As the two conventional algorithms cannot exploit the full state information of the environment and divide the tasks into non-overlapping sets, the resulting UAV trajectories lead to higher normalized accumulated AoIs compared to the CA2C algorithm.

\begin{figure}[!t]
	\center{\includegraphics[width=2.7in]{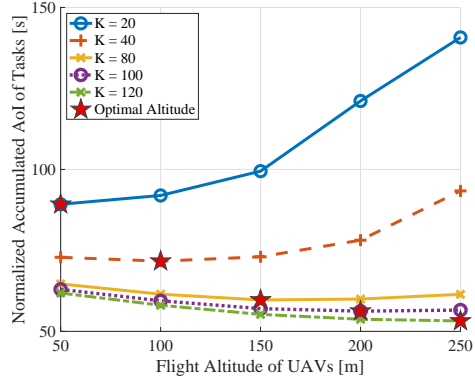}}
	\vspace{-.5em}
	\caption{Normalized accumulated AoI of the tasks versus UAVs' flight altitude for different numbers of subcarriers $K$. The UAVs are trained with the proposed CA2C algorithm. }
	\label{fig: influence of number of sub-carrier and uav flying altitude}
	\vspace{-1em}
\end{figure}

Fig.~\ref{fig: influence of number of sub-carrier and uav flying altitude} shows the normalized accumulated AoI of the tasks versus the UAVs' altitude for different numbers of subcarriers $K$.
As the number of available subcarriers increases, the normalized accumulated AoI of the tasks decreases. 
This is because the transmission of the sensing results takes less time as the number of available subcarriers increases.
For each $K$, there exists an optimal flight altitude for the UAVs, which can be explained as follows.
On the one hand, when the UAVs have a low flight altitude, they have a short sensing range, which results in long flight distances between two successive sensing locations.
On the other hand, when the flight altitude of the UAVs is high, the distance between the UAVs and the BS is long, which results in low data rates and long transmission durations~(i.e., more transmission cycles).
Therefore, there exists an optimal flight altitude for the UAVs for the minimization of the normalized accumulated AoI.
Besides, it can be observed that the optimal flight altitude increases with $K$.
This is because, when $K$ is small, the transmission duration is the dominant factor for the normalized accumulated AoI, and thus, the optimal flight height of the UAVs is low.
However, as $K$ grows larger, the flight duration becomes the dominant factor, which results in larger optimal flight heights.

\begin{figure}[!t] 
\center{\includegraphics[width=0.65\linewidth]{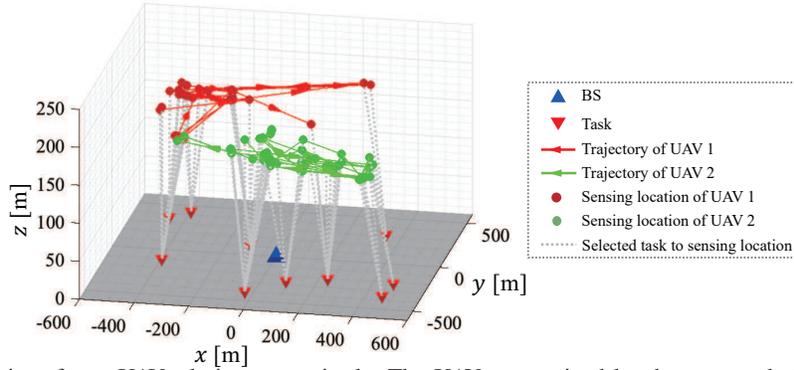}}
	\vspace{-1em}
    \vspace{-1em}
	\caption{ {$3$D trajectories of two UAVs during one episode. 
The UAVs are trained by the proposed algorithm and can select sensing locations with altitudes in the range of $[100,200]$~m.
$K=80$.}}
	\label{fig: 3d trajectory}
	\vspace{-1em}
\end{figure}

Next, we illustrate that the proposed method can also be extended to $3$D distributed trajectory design, where the UAVs can hover within a certain altitude range.
We allow the UAVs to select sensing locations with altitudes in the range of $[100,200]$~m.
To be specific, UAV $i$'s action for the sensing location of Task $T_i$ can be expressed as $\hat{\bm x}_i\in\{ \bm x~|~\|\bm x - \bm x_{T_i}^t\|\cdot \sin\phi \leq h\cdot\tan\phi\}$.
Fig.~\ref{fig: 3d trajectory} shows the trajectories and sensing locations of the UAVs obtained with the proposed CA2C algorithm.
As can be observed, the proposed algorithm has trained the two UAVs to divide the tasks into two non-overlapping sets and to select sensing locations which are nearby, as in Fig.~\ref{fig: cruise trajectory}~(c).
Besides, the average altitude of the UAVs' selected sensing locations is $161$~m, which is close to the optimal altitude of the UAVs when $K=80$, i.e., $150$~m, in Fig.~\ref{fig: influence of number of sub-carrier and uav flying altitude}.

\section{Conclusion}
In this paper, we have investigated the trajectory design problem for the cellular Internet of UAVs, where the UAVs cooperate to execute multiple sensing tasks continuously in order to minimize the AoI of the tasks accumulated over a certain period of time.
To coordinate the multiple UAVs, we have proposed a distributed sense-and-send protocol.
Based on this protocol, we have formulated the trajectory design problem for the cooperative cellular Internet of UAVs as an MDP with the objective to minimize the normalized accumulated AoI of the tasks.
To solve the MDP, we have proposed a CA2C algorithm based on the DQN and DDPG algorithms, which can handle agents with actions involving both discrete and continuous variables.

Our simulation results have led to the following interesting observations. 
 {First, the proposed CA2C algorithm was shown to outperform four benchmark algorithms in terms of AoI minimization.}
Second, cooperative sensing and transmission can considerably reduce the AoI compared to a non-cooperative approach, and the AoI reduction increases with the number of UAVs.
Third, there exists an optimal flying altitude for the UAVs for AoI minimization, which increases with the number of available subcarriers.

\bibliographystyle{IEEEtran}
\bibliography{main}

\end{document}